\documentclass[aip,jcp,reprint,noshowkeys]{revtex4-1}
\usepackage{graphicx,dcolumn,bm,xcolor,microtype,multirow,amscd,amsmath,amssymb,amsfonts,mhchem,longtable,wrapfig}

\usepackage{natbib}
\usepackage[extra]{tipa}
\usepackage{mathpazo,libertine}

\usepackage{comment}
\usepackage{hyperref}
\hypersetup{
    colorlinks=true,
    linkcolor=blue,
    filecolor=blue,
    urlcolor=blue,
    citecolor=blue
}

\definecolor{darkgreen}{HTML}{009900}
\usepackage[normalem]{ulem}

\usepackage{hyperref}
\hypersetup{
    colorlinks=true,
    linkcolor=blue,
    filecolor=blue,
    urlcolor=blue,
    citecolor=blue
}

\newcommand{\UEG}{\text{UEG}}
\newcommand{\HF}{\text{HF}}

\newcommand{\n}[2]{n_{#1}^{#2}}

\newcommand{\rsmu}[2]{\mu_{#1}^{#2}}

\newcommand{\bra}[1]{\ensuremath{\langle #1 \vert}} 
\newcommand{\ket}[1]{\ensuremath{\vert #1  \rangle}} 
\newcommand{\braket}[2]{\ensuremath{\langle  #1 \vert #2  \rangle}}

\newcommand{\Bas}{\mathcal{B}}

\newcommand{\br}[1]{\mathbf{r}_{#1}}

\newcommand{\basis}[0]{\mathcal{B}}
\newcommand{\mub}[0]{\mu^\basis({\bf r})}

\newcommand{\EpsoB}{\mathcal{E}_0^{\basis}}

\newcommand{\LCT}{Laboratoire de Chimie Th\'eorique, UMR 7616, Sorbonne Universit\'e and CNRS, F-75005 Paris, France}
\newcommand{\IUF}{Institut Universitaire de France, F-75005 Paris, France}

\begin{document}	

\title{Accelerated basis-set convergence of coupled-cluster excitation energies using the density-based basis-set correction method}

\author{Diata Traore }
\affiliation{\LCT}
\author{Julien Toulouse}
\affiliation{\LCT}
\affiliation{\IUF}
\author{Emmanuel Giner}
\email[Corresponding author: ]{emmanuel.giner@lct.jussieu.fr}
\affiliation{\LCT}

\date{February 23, 2024}

\begin{abstract}
We present the first application to real molecular systems of the recently proposed linear-response
theory for the density-based basis-set correction method [J. Chem. Phys. {\bf 158}, 234107 (2023)].
We apply this approach to accelerate the basis-set convergence of excitation energies in the 
equation-of-motion coupled-cluster singles doubles (EOM-CCSD) method. We use an approximate linear-response framework which neglects the second-order derivative of the basis-set correction density functional and consists in simply adding to the usual Hamiltonian the one-electron potential generated by the first-order derivative of the functional. This additional basis-set correction potential is evaluated at the Hartree-Fock density, leading to a very computationally cheap basis-set correction. 
We tested this approach over a set of about 30 excitation energies computed for five small molecular systems and found that the excitation energies from the ground state to Rydberg states are the main source of basis-set error. These excitation energies systematically increase when the size of the basis set is increased, suggesting a biased description in favour of the excited state.  
Despite the simplicity of the present approach, the results obtained with the basis-set corrected EOM-CCSD method 
are encouraging as they yield to a mean absolute deviation of 0.02 eV for the aug-cc-pVTZ basis set, while it is of 0.04 eV using the standard EOM-CCSD method. 
This might open the path to an alternative to explicitly correlated approaches to accelerate the basis-set convergence of excitation energies.   
\end{abstract}

\maketitle

\section{Introduction}
One of the main bottleneck of computational electronic-structure wave-function methods is the slow convergence of the 
energy and properties with the size of the one-electron basis set used to expand the wave function. 
At the origin of this limitation lies the divergence of the Coulomb electron-electron interaction as the interelectronic distance goes to zero, creating 
derivative discontinuities in the wave function\cite{Kat-CPAM-57} which cannot be represented with a finite 
one-electron basis set\cite{Hyl-ZP-29}. One can drastically improve the basis-set convergence by using geminals explicitly depending 
on the interelectronic distance\cite{Kut-TCA-85,KutKlo-JCP-91}. 
Among the various flavours of such explicitly correlated approaches, 
the so-called F12 method\cite{NogKut-JCP-94,TenNog-WIREs-12,Ten-TCA-12,HatKloKohTew-CR-12,GruHirOhnTen-JCP-17} 
has proven to be a remarkably efficient tool to describe ground-state properties. 
Applications to excited states
\cite{FliHatKlo-JCP-06, NeiHatKlo-JCP-06, HanKoh-JCP-09, Koh-JCP-09, ShiWer-JCP-10, ShiKniWer-JCP-11, ShiWer-JCP-11, ShiWer-MP-13,HofSchKloKoh-JCP-19} of the F12 or R12 methodology is nevertheless not straightforward 
and the initial attempt led to relatively disappointing results\cite{FliHatKlo-JCP-06}.  
The main issue  came from a strong bias toward the ground state: 
in its usual formulation the geminals are applied only on the Hartree-Fock (HF) Slater determinant which dominates 
the ground-state wave function but has only a small contribution to the excited states.  
Further developments in which the geminals were also applied on singly excited configurations led to 
a strong reduction of this bias\cite{NeiHatKlo-JCP-06}.  
While the latter developments used the so-called linear R12 geminals\cite{FliHatKlo-JCP-06, NeiHatKlo-JCP-06}, 
the modern flavours of explicitly correlated methods use Slater-type geminals (F12) which, unlike the R12 geminals, decay at 
large interelectronic distance and the so-called SP ansatz introduced by Ten-No\cite{Ten-JCP-04} which avoids the optimization of the 
geminal amplitudes by using first- and second-order cusp conditions. 
In this spirit, K\"ohn proposed the so-called XSP ansatz\cite{Koh-JCP-09} which is 
an extended version of the SP ansatz adapted to response theory by adding the single hole-particle 
excitation channels in the SP geminals. The initial application of the XSP ansatz 
to the equation-of-motion coupled-cluster singles doubles (EOM-CCSD) method (see, e.g., Ref.~\onlinecite{StaBar-JCP-93}), in addition to the use of the complementary auxiliary basis-set one-electron correction, suppressed not only the ground-state bias of the standard SP approach 
but also improved the basis-set convergence of the excitation energies. 
Further extensions of the XSP ansatz to the second-order coupled-cluster (CC2) method 
were reported with applications to organic molecules\cite{HofSchKloKoh-JCP-19}. 
The XSP ansatz can therefore be considered as the state-of-the-art method for treating excited states within the F12 framework. 

Despite the undeniable successes of F12 theories, we might point out some limitations. 
First, they are usually formulated for single-reference methods and are not often generally available for multi-reference methods.  Second, they rely on a relatively involved formulation which makes their incorporation in computer software a relatively complex task. Third, the quality of the results for excitation energies particularly depends on the Slater geminal parameter $\gamma$. Last but not least, F12 theories need three- and four-electron integrals which have to be approximated through resolution-of-identity techniques. 

An alternative to F12 approaches was recently proposed by some of the present 
authors\cite{GinPraFerAssSavTou-JCP-18} with the so-called density-based basis-set correction (DBBSC) method. 
This method uses range-separated density-functional theory (RSDFT) 
(see Ref. \onlinecite{TouColSav-PRA-04} and references therein)   
in order to capture the short-range correlation energy missing from the description of wave-function approaches 
with an incomplete one-electron basis set $\basis$. 
The DBBSC method relies on the determination of an effective local range-separation parameter $\mub$, depending on the spatial position $\br{}$,
which provides a local measure of the incompleteness of a given basis set $\basis$. 
One can then simply use a short-range correlation density functional designed 
for multideterminant RSDFT\cite{TouGorSav-TCA-05,FerGinTou-JCP-19} with this local range-separation parameter $\mub$ 
to obtain an estimation of the correlation energy missing in the basis set $\basis$.  
The DBBSC method has been tested and validated for the calculation of ground-state atomization 
energies\cite{LooPraSceTouGin-JPCL-19,GinSceLooTou-JCP-20,YaoGinLiTouUmr-JCP-21,YaoGinAndTouUmr-JCP-21} 
including light and transition-metal elements and strongly correlated systems, 
ionization potentials\cite{GinPraFerAssSavTou-JCP-18,LooPraSceGinTou-JCTC-20}, and 
dipole moments\cite{GinTraPraTou-JCP-21,TraTouGin-JCP-22}. 
Efficient implementations using density-fitting technologies were recently reported\cite{MesKal-JCTC-23,HesGinReiKnoWerTou-JCC-24}. 

Being based on the Levy-Lieb formulation of density-functional theory (DFT), the DBBSC method is a ground-state theory 
and an attempt to apply it to excited states was proposed\cite{GinSceTouLoo-JCP-19} in a state-specific 
flavour by simply evaluating the basis-set correction energy functional at the density (and, in some cases, the pair density) of 
each excited state. 
While the results obtained were encouraging\cite{GinSceTouLoo-JCP-19}, 
this approach is nevertheless not rigorous as it consists in the unjustified
application of a ground-state theory to excited states. 
Recently, the present authors derived the general equations of linear-response theory for the 
DBBSC method\cite{TraGinTou-JCP-23} and applied them to the calculation of excitation energies in a one-dimensional model system. 
This linear-response formalism relies on a variational self-consistent version of the DBBSC method that 
was implemented and tested in Ref. \onlinecite{GinTraPraTou-JCP-21}. 
In the latter framework, the wave function is changed through the addition 
of a basis-set correction potential which is self-consistently determined, and the numerical tests showed that 
while self-consistency makes very little changes on the energy, 
the modified wave function leads to a significant change of the density that greatly 
accelerates the basis-set convergence of dipole moments. 
Nevertheless, the self-consistent approach is difficult to realise with a non-variational wave-function ansatz, which therefore restricts 
its domain of applicability. Recently, the present authors\cite{TraTouGin-JCP-22} proposed to calculate the basis-set correction to coupled-cluster dipole 
moments through numerical energy derivative of the non-self-consistent basis-set correction functional evaluated at the HF density. 
The results showed that the accuracy is similar to that obtained with the self-consistent approach. Such a finding supports the idea that using the self-consistently optimised density in the presence of the basis-set correction is often not mandatory and that one can simply use the HF density in the basis-set correction functional in many cases. 

In the present work, for calculating excitation energies, we propose to use a simple linear-response variant of the DBBSC method which consists 
in several approximations: i) neglecting the second-order derivative of the basis-set correction functional, ii) approximating the density 
at the HF level, iii) using the simplest (semi)local basis-set correction functionals, 
iv) approximating full-configuration-interaction (FCI) excitation energies at the EOM-CCSD level. 
The paper is organised as follows. In Sec.~\ref{sec:theory} we present the theory: 
a summary of the main equations of the ground-state DBBSC method is provided in Sec.~\ref{sec:ground}, then 
the approximate linear-response theory is presented in Sec.~\ref{sec:resp}, and eventually 
its application to the EOM-CCSD method is sketched out in Sec.~\ref{sec:eomcc}. 
Numerical results are presented and discussed in Sec.~\ref{sec:results} for a set of 30 excitation energies in the NH$_3$, H$_2$O, CO, N$_2$ and N$_2$CH$_2$ molecules comprising Rydberg and valence excited states with singlet and triplet symmetry. Finally, Sec.~\ref{sec:conclu} contains our conclusions.

\section{Theory}
\label{sec:theory}
\subsection{Ground-state DBBSC method}
\label{sec:ground}
As introduced in Ref. \onlinecite{GinPraFerAssSavTou-JCP-18}, in the DBBSC method, one defines an approximation to the ground-state energy $E_0^\basis$ for a given basis set $\basis$ by restricting the DFT ground-state energy minimization to $\basis$-representable densities $n^\basis$ (\textit{i.e.} densities that can be obtained from a wave function $\Psi^\basis$ in the many-electron Hilbert space ${\cal H}^\basis$ generated by the basis set $\basis$) 
\begin{equation}
 \label{eq:levy_lieb}
 \begin{aligned}
  E_0^\basis = \min_{n^\basis} \left( F[n^\basis] +\int \text{d}\br{} \, v_\text{ne}(\br{}) n(\br{})\right), 
 \end{aligned}
\end{equation}
where  $F[n] = \min_{\Psi \rightarrow n} \bra{\Psi}\hat{T}+\hat{W}_\text{ee} \ket{\Psi}$ is the usual Levy-Lieb universal density functional. Here, $\hat{T}$ and $\hat{W}_\text{ee}$ are the kinetic-energy and Coulomb electron-electron operators, and $v_\text{ne}$ is the nuclei-electron potential.
As the basis-set restriction in Eq. \eqref{eq:levy_lieb} is only on the density and not on the wave function, the energy $E_0^\basis$ is a much better approximation to the exact ground-state energy $E_0$ than the ground-state FCI energy $E_\text{FCI}^\basis$ in the same basis set $\basis$. 
One can in fact rewrite Eq. \eqref{eq:levy_lieb} in terms of a minimization over wave functions $\Psi^\basis$ restricted to the Hilbert space ${\cal H}^\basis$  
\begin{equation}
 \label{eq:schro_1}
 \begin{aligned}
  E_0^\basis = \min_{\Psi^\basis} \left( \bra{\Psi^\basis} \hat{H} \ket{\Psi^\basis} + \bar{E}^{\basis}[n_{\Psi^\basis}] \right),  
 \end{aligned}
\end{equation}
where $\hat{H}=\hat{T}+\hat{W}_\text{ee} + \hat{V}_\text{ne}$ is the total Hamiltonian, including the nuclei-electron operator $\hat{V}_\text{ne} = \int \text{d} \br{} \, v_\text{ne}(\br{}) \hat{n}(\br{})$ expressed with the density operator $ \hat{n}(\br{})$, and  $\bar{E}^{\basis}[n_{\Psi^\basis}] $ is the basis-set correction density functional evaluated at the density of $\Psi^\basis$. For a $\basis$-representable density $n^\basis$, this functional is defined as 
\begin{eqnarray}
 \Bar{E}^{\basis}[n^\basis] &=& \min_{\Psi \rightarrow n^\basis} \bra{\Psi} \hat{T}+\hat{W}_\text{ee} \ket{\Psi} 
 \nonumber\\
 &&                         - \min_{\Psi^\basis \rightarrow n^\basis} \bra{\Psi^\basis} \hat{T}+\hat{W}_\text{ee} \ket{\Psi^\basis}, 
\end{eqnarray}
and corrects for the error due to the basis-set restriction on the wave functions $\Psi^\basis$. 
A minimizing wave function $\Psi_0^\basis$ in Eq. \eqref{eq:schro_1} satisfies the following self-consistent Schrödinger-like equation 
\begin{equation}
 \label{eq:stat_schro}
 \hat{P}^\basis \hat{\bar{H}}^{\basis}[n_{\Psi_0^\basis}] \ket{\Psi_0^\basis} = \EpsoB \ket{\Psi_0^\basis},
\end{equation}
where $\EpsoB$ is the Lagrange multiplier imposing the normalization condition of the wave function, $\hat{P}^\basis$ is the projector on the Hilbert space ${\cal H}^\basis$, and $\hat{\bar{H}}^{\basis}[n]$ is an effective Hamiltonian, 
\begin{equation}
 \label{eq:hbarb}
 \hat{\bar{H}}^{\basis}[n] =  \hat{H} + \hat{\bar{V}}^{\basis}[n],
\end{equation}
where $\hat{\bar{V}}^{\basis}[n]$ is the basis-set correction potential operator generated by the derivative of the basis-set correction functional 
\begin{equation}
 \label{eq:vbarb}
 \hat{\bar{V}}^{\basis}[n] = \int \text{d}\br{} \frac{\delta \bar{E}^{\basis}[n] }{\delta n(\br{})} \hat{n}(\br{}),
\end{equation}
which corresponds to a one-electron potential. 

Instead of performing the minimization in Eq. \eqref{eq:schro_1}, one can use a non-self-consistent approximation~\cite{GinPraFerAssSavTou-JCP-18,LooPraSceTouGin-JPCL-19}
\begin{equation}
 \label{eq:non_scf}
 E_0^\basis \approx E_{X}^{\basis} + \Bar{E}^{\basis}[n_{Y}^\basis],
\end{equation}
where $E_{X}^{\basis}$ is an approximation of the FCI energy in the basis set $\basis$ calculated with a method $X$ and $n_{Y}^\basis$ is the density calculated with a method $Y$ with the same basis set $\basis$ . This non-self-consistent variant was successfully applied in Refs. \onlinecite{LooPraSceTouGin-JPCL-19,TraTouGin-JCP-22} using coupled cluster singles doubles and perturbative triples [CCSD(T)] for method $X$ and HF for method $Y$ to compute ground-state atomization energies and dipole moments. A detailed study was carried out in Ref. \onlinecite{GinTraPraTou-JCP-21} where it was shown that the non-self-consistent approximation was good enough to calculate the energy.

Unlike the Levy-Lieb density functional, the basis-set correction functional $\bar{E}^{\basis}[n]$ is no longer universal as it depends explicitly on the basis set $\basis$, which is in practice system-dependent. Nevertheless, as most of the basis-set incompleteness consists in missing correlation effects occurring at short interelectronic distances (\textit{i.e.} in the vicinity of the universal electron-electron cusp), one can expect to find generic approximations for $\bar{E}^{\basis}[n]$. As originally proposed in Ref. \onlinecite{GinPraFerAssSavTou-JCP-18}, the basis-set correction functional $\Bar{E}^{\basis}[n]$ can be approximated by 
the short-range (sr) multideterminant (md) correlation functional from RSDFT\cite{TouGorSav-TCA-05,FerGinTou-JCP-19} evaluated with a basis-set dependent and local range-separation parameter $\mu^{\basis}(\br{})$. A semilocal version of it which is appropriate for weakly correlated systems is~\cite{LooPraSceTouGin-JPCL-19}
\begin{equation}
        \label{eq:def_pbe_tot}
        \Bar{E}^{\Bas}[\n{}{}] \approx
        \int \text{d}\br{} \, e_{\text{c,md}}^\text{sr}(\n{}{}(\br{}),\nabla n(\br{}),\rsmu{}{\Bas}(\br{})),
\end{equation}
where $\nabla n$ is the density gradient and $e_{\text{c,md}}^\text{sr}(n,\nabla n,\rsmu{}{})$ is the following correlation energy density
\begin{equation}
        \label{eq:epsilon_cmdpbe}
        e_{\text{c,md}}^\text{sr}(n,\nabla n,\rsmu{}{}) = \frac{e_{\text{c}}{}(n,\nabla n)}{1 + \beta(n,\nabla n) \rsmu{}{3} },
\end{equation}
with
\begin{equation}
        \label{eq:beta_cmdpbe}
        \beta(n,\nabla n) = \frac{3}{2\sqrt{\pi} (1 - \sqrt{2} )} \frac{e_{\text{c}}{}(n,\nabla n)}{\n{2}{\UEG}(\n{}{})},
\end{equation}
where $e_{\text{c}}{}(n,\nabla n)$ can be any approximate Kohn-Sham (semi)local correlation energy density, and 
$\n{2}{\UEG}(n)= \n{}{2} g_0(n)$ is the on-top pair density 
of the uniform electron gas (UEG)\cite{LooGil-WIRES-16} written in terms of the UEG on-top pair-distribution function $g_0(n)$ as parametrised in Eq.~(46) of Ref.~\onlinecite{GorSav-PRA-06}.  
The function $e_{\text{c,md}}^\text{sr}(n,\nabla n,\rsmu{}{})$ 
is designed such that it interpolates between the exact large-$\rsmu{}{}$ 
behavior\cite{TouColSav-PRA-04, GorSav-PRA-06, PazMorGorBac-PRB-06} in $1/\mu^3$ and 
the Kohn-Sham correlation energy density $e_{\text{c}}{}(n,\nabla n)$ at $\rsmu{}{}=0$. 
In the present work, we use two different approximations for the correlation functional $e_{\text{c}}{}(n,\nabla n)$: the local-density approximation (LDA)\cite{PerWan-PRB-92} and the Perdew-Burke-Ernzerhof (PBE) approximation\cite{PerBurErn-PRL-96}.

Although in its most general form $\mu^{\basis}(\br{})$ depends on a correlated wave function $\Psi^\basis_\text{loc}$ used to localise the Coulomb two-electron interaction projected 
in the basis set\cite{GinPraFerAssSavTou-JCP-18}, it was shown in a series of studies
\cite{GinPraFerAssSavTou-JCP-18,LooPraSceTouGin-JPCL-19,LooPraSceGinTou-JCTC-20,GinTraPraTou-JCP-21,TraTouGin-JCP-22,MesKal-JCTC-23,HesGinReiKnoWerTou-JCC-24}
that simply using the HF wave function (\textit{i.e.} $\Psi^\basis_\text{loc}=\Phi_{\text{HF}}^{\basis}$) is enough to obtain reliable results for weakly correlated systems. 
Also, as in most cases wave-function calculations are performed with the frozen-core (FC) approximation, a corresponding FC version for the calculation of $\mu^{\basis}(\br{})$ was introduced in Ref.~\onlinecite{LooPraSceTouGin-JPCL-19}. 
The use of the FC version with the HF wave function leads to the following expression for $\mu^{\basis}(\br{})$
\begin{equation}
 \label{eq:mu_r}
 \mu^{\basis}(\br{}) = \frac{\sqrt{\pi}}{2} \frac{f^{\basis}_\HF(\br{})}{n_{2,\HF}^\basis(\br{})},
\end{equation}
with the function
\begin{equation}
 \label{eq:f_r}
 f^{\basis}_{\HF}(\br{}) = 2 \sum_{p,q\in \text{all}} \sum_{i,j\in\text{act}} \phi_p(\br{}) \phi_q(\br{}) \phi_i(\br{}) \phi_j(\br{}) V_{pq}^{ij},
\end{equation}
and the HF on-top pair density
\begin{equation}
 \label{eq:n2_r}
 n_{2,{\HF}}^{\basis}(\br{}) = 2 \sum_{i,j \in \text{act}} \big( \phi_i(\br{}) \phi_j(\br{})\big)^2 ,
\end{equation}
where $\{\phi_p\}$ are the (real-valued) spatial HF orbitals, $V_{pq}^{ij} = \braket{pq}{ij}$ are the usual two-electron Coulomb integrals, 
and $p,q$ run over all (occupied+virtual) HF spatial orbitals in the basis set $\basis$ and $i,j$ run over only the active (\textit{i.e.} non-core occupied) spatial HF orbitals. 
Correspondingly, with the FC approximation, the basis-set correction functional is evaluated at the active HF density (removing the contribution from the core orbitals)
\begin{equation}
 \label{eq:2_r}
 n_{\HF}^{\basis}(\br{}) = 2 \sum_{i\in\text{act}} \phi_i(\br{})^2.
\end{equation}
As shown in Ref. \onlinecite{GinPraFerAssSavTou-JCP-18}, this local range-separation parameter $\mu^{\basis}(\br{})$ automatically adapts to the basis set 
and tends to infinity in the complete-basis-set (CBS) limit. This makes the basis-set correction functional $\Bar{E}^{\Bas}[\n{}{}]$ in Eq.~(\ref{eq:def_pbe_tot}) correctly vanish in the CBS limit.

\subsection{Approximate linear-response DBBSC method}
\label{sec:resp}
The extension to linear-response theory of the DBBSC method was recently proposed by the present authors in Ref. \onlinecite{TraGinTou-JCP-23}
when using a FCI wave function. 

In this case, the (normalised) ground-state wave function $\Psi_0^\basis$ satisfying Eq.~(\ref{eq:stat_schro}) is expanded in terms of $N$ orthonormal Slater determinants $\{ \Phi_i \}_{1\leq i \leq N}$ spanning the Hilbert space ${\cal H}^\basis$
\begin{equation}
 \ket{\Psi_0^\basis} = \sum_{i=1}^{N} c_{0,i} \ket{\Phi_i}, 
\end{equation}
and the (real-valued) coefficients $\{ c_{0,i} \}$ satisfy the stationary equation, for $1 \leq i \leq N$,
\begin{equation}
\bra{\bar{\Psi}_i} \hat{\bar{H}}^{\basis}[n_{\Psi_0^\basis}] \ket{\Psi_0^\basis} = 0,
\label{eq:stateq}
\end{equation}
where $\bar{\Psi}_i$ are the wave-function derivatives
\begin{equation}
 \ket{\bar{\Psi}_i} = \ket{\Phi_i} - c_{0,i} \ket{\Psi_0^\basis}, 
\end{equation}
which are orthogonal to $\Psi_0^\basis$, \textit{i.e.} $\braket{\bar{\Psi}_i}{\Psi_0^\basis}=0$.

In the linear-response equations of Ref. \onlinecite{TraGinTou-JCP-23}, if we neglect the kernel contribution coming from the second-order derivative of the basis-set correction functional $\bar{E}^\basis[n]$, we obtain the following approximate linear-response equations 
\begin{equation}
 \label{eq:response}
  \bar{\bold{A}} \, \bold{X}_n  = \omega^\basis_n \, \bar{\bold{S}} \, \bold{X}_n, 
\end{equation}
where $\bold{X}_n$ are eigenvectors, $\omega^\basis_n$ are the eigenvalues corresponding to excitation energies ($1\leq n \leq N-1$),  and the matrix elements of $\bar{\bold{A}}$ and $\bar{\bold{S}}$ are, for $2 \leq i,j \leq N$,
\begin{eqnarray}
 \bar{A}_{ij} &=& \bra{\bar{\Psi}_i} \hat{\bar{H}}^{\basis}[n_{\Psi_0^\basis}] - \EpsoB  \ket{\bar{\Psi}_j},
\end{eqnarray}
and
\begin{equation}
\bar{S}_{ij} = \braket{\bar{\Psi}_i}{\bar{\Psi}_j} = \delta_{ij} - c_{0,i} c_{0,j}.
\end{equation}
To avoid the parameter redundancy due to the normalization constraint on the wave function, the first wave-function derivative $\ket{\bar{\Psi}_1} = \ket{\Phi_1} - c_{0,1} \ket{\Psi_0^\basis}$, involving the HF Slater determinant ${\Phi_1} \equiv {\Phi_\text{HF}^\basis}$, has been dropped in the linear-response equations. Thus, it remains only $N-1$ equations.
These approximate linear-response equations are in fact completely equivalent to the FCI equations for the effective Hamiltonian  $\hat{\bar{H}}^{\basis}[n_{\Psi_0^\basis}]$. To see this, we begin by rewriting Eq. \eqref{eq:response} as 
\begin{equation}
 \begin{aligned}
  \bold{\bar{H}} \, \bold{X}_n = \mathcal{E}_n^\basis \, \bar{\bold{S}} \, \bold{X}_n, 
 \end{aligned}
 \label{eq:barHeigenval}
\end{equation}
where $\bar{H}_{ij} = \bra{\bar{\Psi}_i} \hat{\bar{H}}^{\basis}[n_{\Psi_0^\basis}] \ket{\bar{\Psi}_j}$ and $\mathcal{E}_n^\basis = \mathcal{E}_0^\basis + \omega_n^\basis$ are the excited-state total energies. Clearly, Eq.~\eqref{eq:barHeigenval} is the eigenvalue equation for the Hamiltonian $\hat{\bar{H}}^{\basis}[n_{\Psi_0^\basis}]$ in the non-orthogonal basis $\{ \bar{\Psi}_i \}_{2\leq i \leq N}$. We then add the ground-state FCI wave function $\Psi_0^\basis$ to this basis, which, using the stationary equation of Eq.~\eqref{eq:stateq}, leads to the following $N\times N$ eigenvalue equation, for $0\leq n \leq N-1$, 
\begin{equation}
 \label{eq:eq_lr_full}
 \begin{aligned}
 \begin{pmatrix}
 \EpsoB & \bold{0} \\
 \bold{0}             & \bold{\bar{H}} \\
 \end{pmatrix}
 \begin{pmatrix}
 \delta_{0n} \\
 \bold{X}_n \\
 \end{pmatrix}
 = \mathcal{E}_n^\basis
 \begin{pmatrix}
 1 & \bold{0} \\
 \bold{0} & \bar{\bold{S}} \\
 \end{pmatrix}
 \begin{pmatrix}
 \delta_{0n} \\
 \bold{X}_n \\
 \end{pmatrix},
 \end{aligned}
\end{equation}
and the FCI ground state is recovered for $n=0$ with $\bold{X}_0={\bf 0}$. Equation~\eqref{eq:eq_lr_full} is thus the eigenvalue equation for the Hamiltonian $\hat{\bar{H}}^{\basis}[n_{\Psi_0^\basis}]$ in the non-orthogonal basis $\{\Psi_0^\basis \}\cup\{ \bar{\Psi}_i \}_{2\leq i \leq N}$. If we rewrite this eigenvalue equation in the orthonormal basis of the $N$ Slater determinants $\{ \Phi_i \}_{1\leq i \leq N}$, we recover a standard FCI eigenvalue equation for the effective Hamiltonian $\hat{\bar{H}}^{\basis}[n_{\Psi_0^\basis}]$
\begin{equation}
  \bold{H} \bold{c}_n =  \mathcal{E}_n^\basis \bold{c}_n,
\end{equation}
with $H_{ij} = \bra{\Phi_i} \hat{\bar{H}}^{\basis}[n_{\Psi_0^\basis}] \ket{\Phi_j}$ and $\bold{c}_n$ are the eigenvectors.
Equivalently, this eigenvalue equation can be written as
\begin{equation}
 \label{eq:fcibarb}
 \hat{P}^\basis \hat{\bar{H}}^{\basis}[n_{\Psi_0^\basis}] \ket{\Psi_n^\basis} = \mathcal{E}_n^\basis \ket{\Psi_n^\basis},
\end{equation}
and the excited-state wave functions are
\begin{equation}
 \label{}
\ket{\Psi_n^\basis} = \sum_{i=1}^{N} c_{n,i} \ket{\Phi_i}.
\end{equation}

Therefore, in the linear-response DBBSC method, when neglecting the kernel coming from the second-order derivative of the basis-set correction functional, the excitation energies $\omega^\basis_n$ can be obtained directly 
by solving the FCI eigenvalue equation with the effective Hamiltonian $\hat{\bar{H}}^{\basis}[n_{\Psi_0^\basis}] = \hat{H} + \hat{\bar{V}}^{\basis}[n_{\Psi_0^\basis}]$ containing the basis-set correction potential operator $\hat{\bar{V}}^{\basis}[n_{\Psi_0^\basis}]$.

\subsection{Application to the EOM-CCSD method}
\label{sec:eomcc}

Although Eq. \eqref{eq:fcibarb} can a priori be solved using any wave-function method targeting excited states, 
the fact that the basis-set correction potential must be evaluated at the density of the ground-state  wave function $\Psi_0^\basis$ is not convenient.
Indeed, it requires to perform a self-consistent ground-state calculation for obtaining $\Psi_0^\basis$.  
Nevertheless, as shown in previous works\cite{GinPraFerAssSavTou-JCP-18,LooPraSceTouGin-JPCL-19,LooPraSceGinTou-JCTC-20,TraTouGin-JCP-22}, 
good results can be obtained when the density $n_{\Psi_0^\basis}$ is approximated by the HF density $n_{\HF}^{\basis}$.  
Therefore, we use here the following approximation 
\begin{equation}
 \label{eq:approx_v}
 \hat{\Bar{H}}^{\basis}[n_{\Psi_0^\basis}] \approx \hat{\Bar{H}}^{\basis}[n_{\HF}^{\basis}] \equiv \hat{\Bar{H}}^{\basis},
\end{equation}
where we have dropped the explicit dependence on $n_{\HF}^{\basis}$ in the Hamiltonian for the sake of simplicity.

%%%%%%%%%%%%%%%%%%%%%%%%%%%%%%%%%%%%%%%%%%%%%%%%%%%%%%%%%%%%%%%%%%%%%%%%%%%%%%%%%%%%%%%%%%%%%%%%%%%%%%%%%%%%%%%%%%%%%%%%%%%
\begin{table*}
    \centering 
    \caption{NH$_3$ molecule: Standard EOM-CCSD and basis-set corrected EOM-CCSD-LDA and EOM-CCSD-PBE excitation energies (eV) with the AV$X$Z basis sets (with $X=\text{D},\text{T},\text{Q},5$). The letter ``R'' indicates the Rydberg nature of the excited states.  The MAD reported was calculated with all the excitation energies except for the $^1$A$_1$ and $^{3}$A$_1$ excited states for which convergence is not yet reached with the AV5Z basis set. 
   }
    \begin{tabular}{c|rrrr|rr|rr} 
    \hline\hline
                  &\multicolumn{4}{c|}{EOM-CCSD}          & \multicolumn{2}{c|}{EOM-CCSD-LDA} & \multicolumn{2}{c}{EOM-CCSD-PBE}   \\  \hline 
                  & AVDZ   & AVTZ  & AVQZ    & AV5Z     & AVDZ     & AVTZ                 & AVDZ     & AVTZ          \\  %\hline
 $^{1}$A$_2$ (R)  & 6.45   & 6.60  & 6.65    & 6.67     & 6.59     & 6.65                 & 6.58     & 6.65       \\  %\hline
 $^{1}$E$_{\phantom{2}}$     (R)  & 8.02   & 8.15  & 8.19    & 8.20     & 8.16     & 8.20                 & 8.15     & 8.20       \\  %\hline
 $^1$A$_1$   (R)  & 9.65   & 9.33  & 9.13    & 8.93     & 9.77     & 9.39                 & 9.77     & 9.39       \\  %\hline
 $^{3}$A$_2$ (R) & 6.15   & 6.30  & 6.35    & 6.37     & 6.28     & 6.35                 & 6.28     & 6.35       \\  %\hline
 $^{3}$E$_{\phantom{2}}$     (R)  & 7.89   & 8.02  & 8.07    & 8.08     & 8.02     & 8.08                 & 8.02     & 8.08       \\  %\hline
 $^{3}$A$_1$ (R) & 8.45   & 8.70  & 8.60    & 8.48     & 8.98     & 8.76                 & 8.98     & 8.76      \\  %\hline
\hline         
MAD(R)& 0.20    & 0.06  & 0.01    &  -       & 0.07     & 0.01                 & 0.07     & 0.01      \\
\hline\hline
    \end{tabular}
    
    \label{tab:nh3}
\end{table*}
%%%%%%%%%%%%%%%%%%%%%%%%%%%%%%%%%%%%%%%%%%%%%%%%%%%%%%%%%%%%%%%%%%%%%%%%%%%%%%%%%%%%%%%%%%%%%%%%%%%%%%%%%%%%%%%%%%%%%%%%%%%
%%%%%%%%%%%%%%%%%%%%%%%%%%%%%%%%%%%%%%%%%%%%%%%%%%%%%%%%%%%%%%%%%%%%%%%%%%%%%%%%%%%%%%%%%%%%%%%%%%%%%%%%%%%%%%%%%%%%%%%%%%%
\begin{table*}
\caption{H$_2$O molecule: Standard EOM-CCSD and basis-set corrected EOM-CCSD-LDA and EOM-CCSD-PBE excitation energies (eV) with the AV$X$Z basis sets (with $X=\text{D},\text{T},\text{Q},5$). The letter ``R'' indicates the Rydberg nature of the excited states. The MAD for all the six excitation energies is reported.
    }
    \centering  
    \begin{tabular}{c|rrrr|rr|rr} 
    \hline\hline
                   &\multicolumn{4}{c|}{EOM-CCSD}   & \multicolumn{2}{c|}{EOM-CCSD-LDA} & \multicolumn{2}{c}{EOM-CCSD-PBE}   \\  \hline 
                   & AVDZ   & AVTZ  & AVQZ   & AV5Z  & AVDZ          & AVTZ            & AVDZ     & AVTZ          \\  %\hline
$^1$B$_1$ (R)      & 7.45   & 7.60  & 7.66   & 7.68  & 7.63          & 7.67            & 7.62     & 7.67 \\  %\hline
$^1$A$_2$ (R)      & 9.21   & 9.36  & 9.42   & 9.44  & 9.40          & 9.44            & 9.39     & 9.44   \\  %\hline
$^1$A$_1$ (R)      & 9.86   & 9.96  & 10.00  & 10.01 & 10.02         & 10.02           & 10.02    & 10.02  \\  %\hline
$^3$B$_1$ (R)      & 7.04   & 7.20  & 7.28   & 7.30  & 7.22          & 7.28            & 7.21     & 7.27   \\  %\hline
$^3$A$_2$ (R)      & 9.05   & 9.20  & 9.26   & 9.28  & 9.23          & 9.27            & 9.21     & 9.27   \\  %\hline
$^3$A$_1$ (R)      & 9.39   & 9.49  & 9.54   & 9.56  & 9.55          & 9.55            & 9.54     & 9.55   \\  %\hline
\hline         
MAD(R)             & 0.21   & 0.08  & 0.02   & -     & 0.04          & 0.01            & 0.05     & 0.01   \\
\hline\hline
    \end{tabular}
    \label{tab:h2o}
\end{table*}
%%%%%%%%%%%%%%%%%%%%%%%%%%%%%%%%%%%%%%%%%%%%%%%%%%%%%%%%%%%%%%%%%%%%%%%%%%%%%%%%%%%%%%%%%%%%%%%%%%%%%%%%%%%%%%%%%%%%%%%%%%%

We then approximately solve Eq. \eqref{eq:fcibarb}, with the approximation of Eq. \eqref{eq:approx_v}, 
using the EOM-CCSD method (see, e.g., Ref.~\onlinecite{StaBar-JCP-93}) as follows. First, the ground-state wave function is approximated as a coupled-cluster singles doubles (CCSD) ansatz
\begin{equation}
\ket{\Psi_{0}^\basis} = e^{\hat{T}} \ket{\Phi_\HF^\basis},
\end{equation}
with $\hat{T} = \hat{T}_1 + \hat{T}_2$ where $\hat{T}_1$ and $\hat{T}_2$ are the usual single and double excitation operators in the basis set $\basis$.
The single- and double-excitation amplitudes are determined from the ground-state CCSD amplitude equations using the Hamiltonian $\hat{\Bar{H}}^{\basis}$
\begin{equation}
\bra{\Phi_\mu} e^{-\hat{T}}\hat{\Bar{H}}^{\basis} e^{\hat{T}} \ket{\Phi_\HF^\basis} = 0, 
\end{equation}
for all singly and doubly excited Slater determinants $\Phi_\mu$ with respect to $\Phi_\HF^\basis$ (which we will denote by $\mu \in \text{SD})$. 
Note of course that the optimal single- and double-excitation amplitudes are not the same as in standard CCSD since we use the effective Hamiltonian $\hat{\Bar{H}}^{\basis}$.
The corresponding ground-state CCSD energy eigenvalue is
\begin{equation}
\mathcal{E}_{0}^\basis = \bra{\Phi_\HF^\basis} e^{-\hat{T}}\hat{\Bar{H}}^{\basis} e^{\hat{T}} \ket{\Phi_\HF^\basis}. 
\end{equation}
Then, we solve the EOM-CCSD equations with the fixed similarity-transformed Hamiltonian $e^{-\hat{T}}\hat{\Bar{H}}^{\basis}e^{\hat{T}}$
\begin{equation}
\bra{\Phi_\mu} e^{-\hat{T}} ( \hat{\Bar{H}}^{\basis} - \mathcal{E}_{0}^\basis ) e^{\hat{T}} \ket{\Psi_{n}^\basis} = \omega^\basis_n \braket{\Phi_\mu}{\Psi_{n}^\basis}, 
\end{equation}
where $\omega^\basis_n$ are the excitation energies ($n\geq 1$) and the excited-state wave functions are expanded on all single and double excitations
\begin{equation}
 \ket{\Psi_n^\basis} = \sum_{\mu \in \text{SD}} c_{n,\mu} \ket{\Phi_\mu}. 
\end{equation}

In practice, the only change to make in the standard EOM-CCSD algorithm is thus to replace the usual one-electron integrals $h_{pq} = \bra{\phi_p} \hat{T} + \hat{V}_\text{ne} \ket{\phi_q}$ by 
\begin{equation}
\label{eq:new_h}
h_{pq} \to h_{pq} + \bar{v}^{\basis}_{pq},
\end{equation}
where $\bar{v}^{\basis}_{pq} = \bra{\phi_p}\hat{\Bar{V}}^{\basis}[n_{\HF}^{\basis}] \ket{\phi_q}$ are the integrals of the  basis-set correction potential.
Consistently with the FC approximation, the one-electron integral $\bar{v}^{\basis}_{pq}$ is set to zero if  $p$ or $q$ refers to a core orbital.

\section{Numerical results}
\label{sec:results}

%%%%%%%%%%%%%%%%%%%%%%%%%%%%%%%%%%%%%%%%%%%%%%%%%%%%%%%%%%%%%%%%%%%%%%%%%%%%%%%%%%%%%%%%%%%%%%%%%%%%%%%%%%%%%%%%%%%%%%%%%%%
\begin{table*}
    \centering 
    \caption{CO molecule: Standard EOM-CCSD and basis-set corrected EOM-CCSD-LDA and EOM-CCSD-PBE excitation energies (eV) with the AV$X$Z basis sets (with $X=\text{D},\text{T},\text{Q},5$). The letter ``V'' indicates the valence nature of the excited states. The MAD for all the six excitation energies is reported.
    }
    \begin{tabular}{c|rrrr|rr|rr} 
    \hline\hline
                     &\multicolumn{4}{c|}{EOM-CCSD}        & \multicolumn{2}{c|}{EOM-CCSD-LDA} & \multicolumn{2}{c}{EOM-CCSD-PBE}   \\  \hline 
                     & AVDZ    & AVTZ  & AVQZ  & AV5Z    & AVDZ           & AVTZ           & AVDZ     & AVTZ          \\  %\hline
$^1 \Pi$    (V)      & 8.67    & 8.59  & 8.57  & 8.57    & 8.76           & 8.63           & 8.74    & 8.62   \\   %\hline
$^1 \Sigma$ (V)      & 10.10   & 9.99  & 9.99  & 10.00   & 10.15          & 10.01          & 10.15   & 10.01  \\   %\hline
$^1 \Delta$ (V)      & 10.21   & 10.12 &10.13  & 10.13   & 10.26          & 10.15          & 10.26   & 10.14  \\   %\hline
$^3 \Pi$    (V)      & 6.38    & 6.36  & 6.36  & 6.36    & 6.43           & 6.38           & 6.41    & 6.38   \\   %\hline
$^3 \Sigma$ (V)      & 8.34    & 8.34  & 8.37  & 8.39    & 8.39           & 8.37           & 8.39    & 8.37  \\   %\hline
$^3 \Delta$ (V)      & 9.29    & 9.24  & 9.24  & 9.25    & 9.35           & 9.25           & 9.35    & 9.26   \\   %\hline
\hline         
MAD(V)               & 0.06    & 0.02  & 0.01  &         & 0.11           & 0.02           & 0.10    &  0.02  \\
\hline\hline
    \end{tabular}                                                                          
    
    \label{tab:co}
\end{table*}
%%%%%%%%%%%%%%%%%%%%%%%%%%%%%%%%%%%%%%%%%%%%%%%%%%%%%%%%%%%%%%%%%%%%%%%%%%%%%%%%%%%%%%%%%%%%%%%%%%%%%%%%%%%%%%%%%%%%%%%%%%%

%%%%%%%%%%%%%%%%%%%%%%%%%%%%%%%%%%%%%%%%%%%%%%%%%%%%%%%%%%%%%%%%%%%%%%%%%%%%%%%%%%%%%%%%%%%%%%%%%%%%%%%%%%%%%%%%%%%%%%%%%%%
\begin{table*}
 \caption{N$_2$ molecule: Standard EOM-CCSD and basis-set corrected EOM-CCSD-LDA and EOM-CCSD-PBE excitation energies (eV) with the AV$X$Z basis sets (with $X=\text{D},\text{T},\text{Q},5$). The letter ``V'' indicates the valence nature of the excited states. The MAD for all the six excitation energies is reported.
    }
    \centering  
    \begin{tabular}{c|rrrr|rr|rr} 
    \hline\hline
                     &\multicolumn{4}{c|}{EOM-CCSD}         & \multicolumn{2}{c|}{EOM-CCSD-LDA} & \multicolumn{2}{c}{EOM-CCSD-PBE}  \\  \hline 
                     & AVDZ  & AVTZ    & AVQZ   & AV5Z    & AVDZ         & AVTZ             & AVDZ             & AVTZ                  \\ % %\hline
$^1 \Pi_\text{g}$      (V)  & 9.50  & 9.41    & 9.40   & 9.40    & 9.60         & 9.45             & 9.57             & 9.45                  \\ %\hline
$^1 \Sigma_\text{u}$ (V)  & 10.20 & 10.00   & 9.98   & 9.98    & 10.24        & 10.01            & 10.24            & 10.01                 \\ %\hline
$^1 \Delta_\text{u}$   (V)  & 10.61 & 10.44   & 10.42  & 10.42   & 10.66        & 10.46            & 10.66            & 10.46                 \\ %\hline
$^3 \Sigma_\text{u}$ (V)  & 7.69  & 7.66    & 7.69   & 7.69    & 7.71         & 7.67             & 7.71             & 7.67                  \\ %\hline
$^3 \Pi_\text{g}$    (V)  & 8.12  & 8.09    & 8.10   & 8.10    & 8.20         & 8.13             & 8.18             & 8.13                  \\ %\hline
$^3 \Delta_\text{u}$ (V)  & 9.04  & 8.91    & 8.91   & 8.91    & 9.07         & 8.92             & 9.07             & 8.92                  \\ %\hline
\hline         
MAD(V)               & 0.11  & 0.01    & 0.00    & -       & 0.16         & 0.03             & 0.15             & 0.03                  \\
\hline\hline
    \end{tabular}
   
    \label{tab:n2}
\end{table*}
%%%%%%%%%%%%%%%%%%%%%%%%%%%%%%%%%%%%%%%%%%%%%%%%%%%%%%%%%%%%%%%%%%%%%%%%%%%%%%%%%%%%%%%%%%%%%%%%%%%%%%%%%%%%%%%%%%%%%%%%%%%
\subsection{Computational details}
We computed systematically the first three excited states of both singlet and triplet spin symmetry 
of the NH$_3$, H$_2$O, CO, N$_2$, and N$_2$CH$_2$ molecules whose geometries have been taken from 
Ref. \onlinecite{LooSceBloGarCafJac-JCTC-18}. This constitutes a set of 30 excited states among which 14 have a Rydberg character and 16 have a valence character (according to the classification 
reported in previous works\cite{StaBar-JCP-93,LooSceBloGarCafJac-JCTC-18}).  

Standard EOM-CCSD calculations (\textit{i.e.} using the standard Hamiltonian) have been performed with the 
Gaussian-16 software\cite{g16} with the aug-cc-pV$X$Z ($X=\text{D},\text{T},\text{Q},5$) basis sets~\cite{KenDunHar-JCP-92}, abbreviated as AV$X$Z, except for the N$_2$CH$_2$ molecule with the AV5Z basis set for which the PySCF software\cite{PySCF} was used. 
The EOM-CCSD calculations using the effective Hamiltonian $\hat{\Bar{H}}^{\basis}$
have been performed using the PySCF software\cite{PySCF} by reading the one- and two-electron 
integrals defining the Hamiltonian $\hat{\Bar{H}}^{\basis}$ 
from a FCIDUMP format for the AVDZ and AVTZ basis sets. 
Limitations of the FCIDUMP format prevented us to perform calculations in larger basis sets, 
but we believe that the results presented here are sufficient to discuss the main trends. 
The one-electron integrals, including the basis-set correction potential integrals $\bar{v}_{pq}^\basis$ [Eq.~\eqref{eq:new_h}], and the two-electron integrals  
have been computed with the Quantum Package software\cite{QP2}. 
The integrals $\bar{v}_{pq}^\basis$ have been computed using a standard Becke-type~\cite{Bec-JCP-88b} spatial 
grid with 75 radial points and 302 Lebedev angular points.  
All calculations have been performed within the FC approximation, 
both for the EOM-CCSD part and the computation of all quantities related to the basis-set correction
[see Eqs. \eqref{eq:mu_r}-\eqref{eq:2_r}]. 
The EOM-CCSD calculations using either the LDA and PBE versions of the basis-set correction functional 
[see Eqs. \eqref{eq:epsilon_cmdpbe} and \eqref{eq:beta_cmdpbe}] will be referred to as EOM-CCSD-LDA and EOM-CCSD-PBE, respectively.

\subsection{Results and discussion}
%%%%%%%%%%%%%%%%%%%%%%%%%%%%%%%%%%%%%%%%%%%%%%%%%%%%%%%%%%%%%%%%%%%%%%%%%%%%%%%%%%%%%%%%%%%%%%%%%%%%%%%%%%%%%%%%%%%%%%%%%%%
\begin{table*}
 \caption{N$_2$CH$_2$ molecule: Standard EOM-CCSD and basis-set corrected EOM-CCSD-LDA and EOM-CCSD-PBE excitation energies (eV) with the AV$X$Z basis sets (with $X=\text{D},\text{T},\text{Q},5$). The letter ``R'' or ``V'' indicates the Rydberg or valence nature of the excited states. The MAD for all the six excitation energies is reported, together with the MAD calculated with the two Rydberg excited states and the MAD calculated with the four valence excited states. 
    } 
    \centering  
    \begin{tabular}{c|rrrr|rr|rr} 
    \hline\hline
                  &\multicolumn{4}{c|}{EOM-CCSD}         & \multicolumn{2}{c|}{EOM-CCSD-LDA} & \multicolumn{2}{c}{EOM-CCSD-PBE}  \\  \hline 
                  & AVDZ  & AVTZ    & AVQZ   & AV5Z    & AVDZ         & AVTZ             & AVDZ             & AVTZ          \\ % %\hline
$^1$A$_2$ (V)     & 3.23  & 3.19    & 3.19   & 3.20    & 3.26         & 3.20             & 3.25             & 3.20          \\ %\hline
$^1$B$_1$ (R)     & 5.43  & 5.57    & 5.62   & 5.65    & 5.58         & 5.63             & 5.58             & 5.63          \\ %\hline
$^1$A$_1$ (V)     & 5.90  & 5.94    & 5.96   & 5.97    & 5.99         & 5.98             & 5.99             & 5.98          \\ %\hline
$^3$A$_2$ (V)     & 2.90  & 2.88    & 2.88   & 2.90    & 2.94         & 2.89             & 2.93             & 2.89          \\ %\hline
$^3$B$_1$ (V)     & 3.99  & 3.95    & 3.95   & 3.95    & 3.99         & 3.95             & 3.98             & 3.94          \\ %\hline
$^3$A$_1$ (R)     & 5.26  & 5.42    & 5.46   & 5.50    & 5.41         & 5.48             & 5.41             & 5.48          \\ %\hline
\hline         
MAD               & 0.10   & 0.04    & 0.02   & -       & 0.05         & 0.01             & 0.05             & 0.01          \\
MAD(R)            & 0.23  & 0.08    & 0.04   & -       & 0.08         & 0.02             & 0.08             & 0.02          \\
MAD(V)            & 0.03  & 0.01    & 0.01   & -       & 0.04         & 0.01              & 0.03             & 0.01          \\
\hline\hline
    \end{tabular}
   
    \label{tab:diazo}
\end{table*}
%%%%%%%%%%%%%%%%%%%%%%%%%%%%%%%%%%%%%%%%%%%%%%%%%%%%%%%%%%%%%%%%%%%%%%%%%%%%%%%%%%%%%%%%%%%%%%%%%%%%%%%%%%%%%%%%%%%%%%%%%%%
We report the EOM-CCSD, EOM-CCSD-LDA, and EOM-CCSD-PBE excitation energies
for the NH$_3$, H$_2$O, CO, N$_2$, and N$_2$CH$_2$ molecules in Tables \ref{tab:nh3}, \ref{tab:h2o}, \ref{tab:co}, \ref{tab:n2}, and \ref{tab:diazo}, respectively.  
We can notice that, except for the $^1$A$_1$ and $^{3}$A$_1$ states of the NH$_3$ molecule, all standard EOM-CCSD excitation energies computed with the AV5Z basis set can be 
considered as converged with respect to the basis set within less than 0.02 eV. 
For each system and each excited state, we will therefore use the EOM-CCSD excitation energies computed with the AV5Z basis set as our estimate for the CBS limit. 
For each system we also report the mean absolute deviation (MAD) with respect to the reference AV5Z basis set calculation, and when possible we also report the 
MAD computed with only Rydberg excited states [MAD(R)] or only valence excited states [MAD(V)] in order to differentiate these two types of excitation energies. 
We also report in Table \ref{tab:mad_tot} the MAD computed over the whole set of 28 converged excitation energies, together with MAD(R) and MAD(V) obtained 
with the 12 and 16 converged Rydberg and valence excitation energies, respectively. 

A detailed look at all the tables reveals two interesting general trends: 
i) except for the $^3 \Sigma$ valence state of the CO molecule 
and for the two unconverged  $^1$A$_1$ and $^{3}$A$_1$ Rydberg states of the NH$_3$ molecule, all the excitation energies corresponding to Rydberg states increase when the size of the basis set is increased
while the excitation energies corresponding valence excited states tend to be stable or decrease with the basis set, 
and ii) the basis-set error in the excitation energies of the Rydberg excited states is much larger than that of the valence excited states. 
Quantitatively, for the Rydberg excited states, the overall MAD is 0.21 eV, 0.07 eV, and 0.02 eV for standard EOM-CCSD with the AVDZ, AVTZ, and AVQZ basis sets, while for the valence excited states it is 0.07 eV, 0.01 eV, and 0.01 eV with the same basis sets. 
A qualitative explanation of this observation could be that, in a Rydberg excited state, one electron is in a diffuse orbital relatively 
far from the bulk of the electronic density of the molecule, and therefore the correlation effects of this electron are much smaller than in the ground state, 
leading, in a small basis set, to a description biased toward the excited state and therefore a too small excitation energy. 
By contrast, in a valence excited state, the excited electron remains in a valence orbital and is much closer to the bulk of the electronic density, 
and therefore the correlation effects are much more comparable to those of the ground state, leading to a much smaller basis-set error. 
We therefore conclude from this part of the study that the main source of basis-set error for the description of a set of excited states come from the Rydberg excited states.

Moving now to the our EOM-CCSD-LDA and EOM-CCSD-PBE calculations, we see that, with respect to the standard EOM-CCSD calculations in a given basis set, 
the effect of the basis-set correction potential is always to increase the excitation energies. 
As the valence excitation energies tend to decrease with the basis set, our approach cannot improve these excitation energies while it will 
improve the description of the Rydberg excitation energies which tend to be underestimated in a finite basis set. 
The present test set consists of 12 Rydberg excited states and 16 excited valence states, and represents a relatively balanced selection between 
excitation energies that the basis-set correction method with the current approximations can improve and excitation energies that it 
will tend to deteriorate. 

From a quantitative point of view, the EOM-CCSD-LDA and EOM-CCSD-PBE approximations give very similar results with the AVDZ basis set 
(the larger difference of the MAD is  0.01 eV) and essentially indistinguishable results with the AVTZ basis set. 
We also notice that the MADs for the Rydberg excitation energies is drastically reduced by the basis-set correction potential. 
With the AVDZ basis set, the MADs obtained with EOM-CCSD-PBE are 0.07 eV, 0.05 eV, and 0.08 eV for the NH$_3$, H$_2$O, and N$_2$CH$_2$ molecules, respectively, 
smaller by at least a factor of two with respect to the MADs obtained with the standard EOM-CCSD method, and thus reaching an accuracy similar to standard EOM-CCSD with the AVTZ basis set. 
With the AVTZ basis set, the MADs obtained the basis-set corrected EOM-CCSD method for the Rydberg excitation energies are 0.01 eV for both the NH$_3$ and H$_2$O molecules, 
and 0.02 eV for the N$_2$CH$_2$ molecule, which is as accurate as standard EOM-CCSD with the AVQZ basis set for the NH$_3$ molecule, 
and even more accurate in the case of the H$_2$O and N$_2$CH$_2$ molecules. 
We therefore conclude that the addition of the basis-set correction potential drastically improves the basis-set convergence of the excitation energies for the Rydberg excited states  
at virtually no cost with respect to standard EOM-CCSD calculations. 

Turning now to the set of valence excited states, as anticipated above, the basis-set correction potential overall deteriorates the accuracy of the excitation energies, 
but this deterioration becomes smaller as the basis-set size increases. 
More quantitatively, the MADs obtained with EOM-CCSD-PBE with the AVDZ basis set are 0.10 eV, 0.15 eV, and 0.03 eV for the CO, N$_2$, and N$_2$CH$_2$ molecules, respectively, 
which are larger than the standard EOM-CCSD values by 0.04 eV for the CO and N$_2$ molecules, but identical for the N$_2$CH$_2$ molecule. With the AVTZ basis set, the MADs obtained with EOM-CCSD-PBE decrease to 0.02 eV, 0.03 eV, and 0.01 eV, representing an accuracy comparable to that of standard EOM-CCSD with the same basis set.

%%%%%%%%%%%%%%%%%%%%%%%%%%%%%%%%%%%%%%%%%%%%%%%%%%%%%%%%%%%%%%%%%%%%%%%%%%%%%%%
\begin{table}
 \caption{Total MAD calculated over the set of 28 excitation energies, 
   together with the MADs calculated over the set of Rydberg excited states [MAD(R)] and the set of valence excited states [MAD(V)].
   }
    \centering  
    \begin{tabular}{l|rrr}
    \hline\hline
                 & MAD         & MAD(R)          & MAD(V) \\
  \hline         
                 &\multicolumn{3}{c }{EOM-CCSD}                 \\
  AVDZ           & 0.14        &  0.21           & 0.07       \\
  AVTZ           & 0.04        &  0.07           & 0.01       \\
  AVQZ           & 0.01        &  0.02           & 0.01       \\
  \hline                       
                 &\multicolumn{3}{c }{EOM-CCSD-LDA}         \\
  AVDZ           & 0.09        &  0.06           & 0.10       \\
  AVTZ           & 0.02        &  0.01           & 0.02       \\
  \hline                       
                 &\multicolumn{3}{c }{EOM-CCSD-PBE}         \\
  AVDZ           & 0.08        &  0.07           & 0.09  \\
  AVTZ           & 0.02        &  0.01           & 0.02  \\
  \hline\hline
    \end{tabular}
   
    \label{tab:mad_tot}
\end{table}
%%%%%%%%%%%%%%%%%%%%%%%%%%%%%%%%%%%%%%%%%%%%%%%%%%%%%%%%%%%%%%%%%%%%%%%%%%%%%%%%%%%%%
\section{Conclusion}
\label{sec:conclu}
In the present work we proposed and tested a novel scheme based on the DBBSC method\cite{GinPraFerAssSavTou-JCP-18} 
to improve the basis-set convergence of the excitation energies in wave-function calculations. 
This is based on the recently introduced linear-response theory\cite{TraGinTou-JCP-23} for the DBBSC method which was only tested 
on a one-dimensional model system. 
In order to treat real molecular systems, 
we use basis-set correction density functionals previously developed in the DBBSC method for ground-state calculations, 
and we introduce a further approximation in the response equations which consists in neglecting the kernel contribution coming from the second-order derivatives of the basis-set correction functional. 
The advantage of the approximation is that the response equations can be rewritten as a usual FCI eigenvalue equation with an additional 
one-electron potential coming from the first-order derivative of the basis-set correction functional. Therefore, by simply modifying the one-electron integrals in the Hamiltonian, this scheme can be applied to any wave-function methods targeting excited states. 

Applying this scheme to the EOM-CCSD method, we presented numerical tests performed on a set of 30 excitation energies on the NH$_3$, H$_2$O, CO, N$_2$, and N$_2$CH$_2$ molecules. 
The results were analysed based on a partition of the excitation energies: the ones corresponding to Rydberg excited states and the ones corresponding to valence excited states. 
We found that the global basis-set error is dominated by the Rydberg excited states, whose excitation energies tend to increase when the size of the basis set is increased, 
while the valence excitation energies tend to be much less sensible to basis set and overall tend to decrease with the basis set. 
The increase of the excitation energy with the basis set for a Rydberg excited state can be qualitatively understood by the fact that the excited electron is relatively far
from the molecule and is therefore much less correlated than in the ground state, 
which leads to a biased description in favour of the excited state in small basis sets. By contrast, in a valence excited state, the excited electron remains in the valence region and 
its correlation effects are much more comparable to that of the ground state, leading to a much smaller sensitivity to finite basis-set effects. 
We found that the present basis-set corrected EOM-CCSD method always increases the excitation energies, and therefore it tends to improve Rydberg excitation energies while it tends to deteriorate 
the valence excitation energies. Indeed, with the AVTZ basis set, the DBBSC scheme reduces the MAD on the Rydberg excitation energies obtained with standard EOM-CCSD from 0.07 eV to 0.01 eV, which is 
a large improvement. With the same AVTZ basis set, the DBBSC scheme increases the MAD on the valence excitation energies from 0.01 eV to 0.02 eV, which still represents a reasonable accuracy. 

We therefore conclude that the present basis-set corrected EOM-CCSD method allows one to overall reduce the basis-set error for the computation of excitation energies, at virtually no additional cost with respect to standard EOM-CCSD calculations. In forthcoming works, we will investigate the impact of taking into account the basis-set correction kernel, together with the dependency on the density used.

\providecommand{\latin}[1]{#1}
\makeatletter
\providecommand{\doi}
  {\begingroup\let\do\@makeother\dospecials
  \catcode`\{=1 \catcode`\}=2 \doi@aux}
\providecommand{\doi@aux}[1]{\endgroup\texttt{#1}}
\makeatother
\providecommand*\mcitethebibliography{\thebibliography}
\csname @ifundefined\endcsname{endmcitethebibliography}
  {\let\endmcitethebibliography\endthebibliography}{}

\end{document}